\documentclass{article}%
\usepackage{amsfonts}
\usepackage{amsmath}
\usepackage{amssymb}
\usepackage{graphicx}%
\begin{document}

\title{Dynamical group approach to conformal field theory}
\author{G. A. Kerimov\\Physics Department, Trakya University, Edirne, Turkey\\E-mail: gkerimov@trakya.edu.tr}
\date{02 July 2025}
\maketitle

\section{Introduction}

The key role of symmetry in modern physics is widely recognized. In the late
1990s, we proposed a group-theoretical approach \cite{ker98} to describe
$S$-matrices of quantum mechanical systems with dynamical symmetries. The
$S$-matrices for all these problems are related to the intertwining operators
between equivalent representations of the underlining dynamical groups ( see,
e.g. \cite{ker02}, \cite{ker05} ). This approach naturally generalizes to
quantum field theories with dynamical symmetry. It can be done by using the
Heisenberg picture in the definition of the $S$-matrix.

In the Heisenberg picture \cite{wei95}, \cite{str13} the $S$-matrix is given by%

\begin{equation}
S_{\alpha_{\beta}}=\left(  \Psi_{\alpha}^{-},\Psi_{\beta}^{+}\right)  =\left(
\Psi_{\alpha}^{-},S\Psi_{\beta}^{-}\right)  \tag{1.1}%
\end{equation}
where $\Psi_{\beta}^{+}$ and $\Psi_{\alpha}^{-}$ are the time-independent `in'
and `out' states, and Greek letters $\alpha$ and $\beta$ are abbreviations
indicating the particle content. The Hilbert spaces $\mathcal{H}^{in}$ and
$\mathcal{H}^{out}$ spanned by the `in' and `out' states are Fock spaces.
Since $\mathcal{H}^{in}=\mathcal{H}^{out}$, the Heisenberg $S$-matrix operator
$S$ is a unitary operator between $\mathcal{H}^{in}$ and $\mathcal{H}^{out}$%
\begin{equation}
S\colon\quad\mathcal{H}^{out}\longrightarrow\mathcal{H}^{in} \tag{1.2}%
\end{equation}
Let $G$ be a dynamical group containing the Poincar\'{e} group $\mathcal{P}$,
the kinematical group of Minkowski spacetime. By analogy with the
non-relativistic case, we assume that the representations of the group $G$
acting on the spaces $\mathcal{H}^{in}$ and $\mathcal{H}^{out}$ are not the
same. Namely, the unitary representations $T^{in}\left(  g\right)  $ and
$T^{out}\left(  g\right)  $ of the group $G$, acting in the spaces
$\mathcal{H}^{in}$ and $\mathcal{H}^{out}$, are induced by equivalent
irreducible representations of the group $G$. Thus, the $S$-matrix is an
intertwining operator between the representations $T^{in}\left(  g\right)  $
and $T^{out}\left(  g\right)  $ of the dynamical group $G$%

\begin{equation}
ST^{out}\left(  g\right)  =T^{in}\left(  g\right)  S,\quad\forall g\in G
\tag{1.3}%
\end{equation}
or%
\begin{equation}
SdT^{out}\left(  X\right)  =dT^{in}\left(  X\right)  S,\quad\forall
X\in\mathfrak{g,} \tag{1.4}%
\end{equation}
where $dT^{in}$ and $dT^{out}$ are the corresponding representations of the
algebra $\mathfrak{g}$ of $G$.

It should be emphasized that \cite{wei95} the same representation $U\left(
\Lambda,\mathbf{a}\right)  $ of the Poincar\'{e} group $\mathcal{P}$ acts on
both spaces $\mathcal{H}^{in}$ and $\mathcal{H}^{out}$. Hence,%
\begin{equation}
SU\left(  \Lambda,\mathbf{a}\right)  =U\left(  \Lambda,\mathbf{a}\right)
S,\quad\left(  \Lambda,\mathbf{a}\right)  \in\mathcal{P} \tag{1.5}%
\end{equation}
for arbitrary Lorentz transformations $\Lambda$ and translations $\mathbf{a}
$. This means that the $S$-matrix is Poincare invariant.

In this paper we discuss the application of this approach to Euclidean quantum
field theories whose dynamics are governed by the Euclidean conformal group.
The objects considered in Euclidean quantum field theory \cite{sch58}\ are not
scattering amplitudes, but the Schwinger functions or Euclidean Green's
functions. As we shall see, the kernel of the intertwining operator is related
to the Schwinger function. In particular, the 2-point Schwinger function is
the kernel of the intertwining operator between equivalent irreducible
representations of the Euclidean conformal group, as already pointed out by
Koller \cite{kol75}.

The remaining of this paper is arranged as follows. In Sec. 1 we present basic
facts about the Euclidean conformal group and notations necessary for
subsequent sections. In Sec. 2 we consider d-dimensional Euclidean conformal
field theory for scalar bosons. The intertwining property of $S$ leads to
powerful constraints on the n-point Schwinger functions, which allows the
definition of two- and three-point functions. Moreover, the Clebsch-Gordan
decomposition for the Euclidean conformal group can be used to define n-point
Schwinger functions in terms of Clebsch-Gordan kernels. For simplicity of
description, in Sec. 3 we consider only four-point functions in two
dimensions. The appendix presents Naimark's \cite{naim1} results on the
decomposition of tensor products of scalar principal series representations,
which are necessary for evaluating four-point functions.

\section{The group $SO_{0}\left(  d+1,1\right)  $ and its Class 1
representations}

In this section we give, mainly to establish notation, a brief review of class
1 (scalar) representations \cite{vkl93}\ of the group $SO_{0}\left(
d+1,1\right)  $.

\subsection{The group $SO_{0}\left(  d+1,1\right)  $ and its subgroups}

Let $R^{d+1,1}$be a $\left(  d+2\right)  $ - dimensional real linear space
with bilinear form%
\begin{equation}
\left[  \xi,\eta\right]  =-\xi_{1}\eta_{1}-\xi_{2}\eta_{2}-...-\xi_{d+1}%
\eta_{d+1}+\xi_{d+2}\eta_{d+2}. \tag{2.1}%
\end{equation}
The upper sheet of the cone $\left[  \xi,\xi\right]  =0,\,\xi_{d+2}>0$ will be
denoted by $C_{+}^{d+1}$. By $SO_{0}\left(  d+1,1\right)  $ we denote the
group of all linear transformations $g$ on $R^{d+1,1}$ that preserve the
bilinear form $\left[  \xi,\eta\right]  $ and for which%
\begin{equation}
\det g=1,\qquad g_{d+2\;d+2}\geq1. \tag{2.2}%
\end{equation}
The Euclidean conformal group is isomorphic to $SO_{0}\left(  d+1,1\right)  $.

Denoting by $N,\overline{N},A,M$ the subgroups of the group $SO_{0}\left(
d+1,1\right)  $, consisting, respectively, of the matrices%
\begin{align}
n\left(  \mathbf{b}\right)   &  =\left(
\begin{array}
[c]{ccc}%
I_{d} & -\mathbf{b}^{t} & \mathbf{b}^{t}\\
\mathbf{b} & 1-\frac{\mathbf{b}^{2}}{2} & \frac{\mathbf{b}^{2}}{2}\\
\mathbf{b} & -\frac{\mathbf{b}^{2}}{2} & 1+\frac{\mathbf{b}^{2}}{2}%
\end{array}
\right)  \,,\quad\overline{n}\left(  \mathbf{c}\right)  =\left(
\begin{array}
[c]{ccc}%
I_{d} & \mathbf{c}^{t} & \mathbf{c}^{t}\\
-\mathbf{c} & 1-\frac{\mathbf{c}^{2}}{2} & -\frac{\mathbf{c}^{2}}{2}\\
\mathbf{c} & \frac{\mathbf{c}^{2}}{2} & 1+\frac{\mathbf{c}^{2}}{2}%
\end{array}
\right)  ,\,\nonumber\\
& \tag{2.3}\\
a_{\lambda}  &  =\left(
\begin{array}
[c]{ccc}%
I_{d} & 0 & 0\\
0 & \frac{\lambda^{2}+1}{2\lambda} & \frac{\lambda^{2}-1}{2\lambda}\\
0 & \frac{\lambda^{2}-1}{2\lambda} & \frac{\lambda^{2}+1}{2\lambda}%
\end{array}
\,\right)  ,\qquad m=\left(
\begin{array}
[c]{ccc}%
m & 0 & 0\\
0 & 1 & 0\\
0 & 0 & 1
\end{array}
\right)  ,\,m\in SO\left(  d\right)  ,\quad\nonumber
\end{align}
we have%
\begin{equation}
SO_{0}\left(  d+1,1\right)  =NAM\overline{N} \tag{2.4}\label{ga}%
\end{equation}
(the equality holds almost everywhere). Here $I_{d}$ is a $d$-dimensional
identity matrix, $\mathbf{b,c}$ are $d$-dimensional row vectors, and
$\mathbf{b}^{t}$, $\mathbf{c}^{t}$ are the corresponding column vectors. We
denote for simplicity an element of $M$ and the corresponding element of
$SO\left(  d\right)  $ by the same letter. The decomposition (\ref{ga}) is
called the Gauss decomposition of $SO_{0}\left(  d+1,1\right)  $. It is worth
noting that $A$ is a dilation subgroup, the subgroup $\overline{N}$ consists
of special conformal transformations, and the subgroup $MN=NM$ is isomorphic
to the inhomogeneous rotation group $ISO(d)$ (Euclidean group). The
isomorphism $MN\cong ISO(d)$ is given by%
\begin{equation}
mn\left(  \mathbf{b}\right)  \,\longrightarrow\,\left(  m,\mathbf{b}\right)
\tag{2.5}%
\end{equation}
where $\left(  m,\mathbf{b}\right)  $ is an element of $ISO(d)$.

\subsection{Class 1 representations of $SO_{0}\left(  d+1,1\right)  $}

Class 1 representations $T^{\sigma}\left(  g\right)  $ of the group
$SO_{0}\left(  d+1,1\right)  $ can be realized in the space of smooth
functions on $C_{+}^{d+1}$such that%
\begin{equation}
F\left(  \omega\xi\right)  =\omega^{\sigma}F\left(  \xi\right)  ,\quad
\omega>0,\quad\xi\in C_{+}^{d+1}, \tag{2.6}%
\end{equation}
where $\sigma$ is an arbitrary complex number. This representations are given
by%
\begin{equation}
\left[  T^{\sigma}\left(  g\right)  F\right]  \left(  \xi\right)  =F\left(
g^{-1}\xi\right)  . \tag{2.7}%
\end{equation}
where $g\in SO_{0}\left(  d+1,1\right)  $.

Almost every $\xi\in C_{+}^{d+1}$ can be represented as%
\begin{equation}
\xi=\omega\zeta,\quad\zeta=\left(  \mathbf{x,}\frac{1}{2}\left(
-1+\mathbf{x}^{2}\right)  ,\frac{1}{2}\left(  1+\mathbf{x}^{2}\right)
\right)  ,\tag{2.8}%
\end{equation}
where $\omega>0$, $\mathbf{x}\in R^{d}$. Consequently, the class 1
representations of $SO_{0}\left(  d+1,1\right)  $\ can be realized on the
space $\mathcal{C}^{\infty}\left(  R^{d}\right)  $ of smooth functions on
$R^{d}$. In this realization the representations $T^{\sigma}\left(  g\right)
$\ of $SO_{0}\left(  d+1,1\right)  $ are given by%
\begin{equation}
\left[  T^{\sigma}\left(  g\right)  f\right]  \left(  \mathbf{x}\right)
=\omega_{g}\left(  \mathbf{x}\right)  ^{\sigma}f\left(  \mathbf{x}_{g}\right)
,\tag{2.9}%
\end{equation}
where $\omega_{g}\left(  \mathbf{x}\right)  $ and $\mathbf{x}_{g}$ are defined
from%
\begin{equation}
g^{-1}\zeta=\omega_{g}\left(  \mathbf{x}\right)  \zeta_{g},\tag{2.10}%
\end{equation}
with%
\begin{equation}
\omega_{g}\left(  \mathbf{x}\right)  =\omega\left(  \mathbf{x,}g\right)
,\quad\zeta_{g}=\left(  \mathbf{x}_{g}\mathbf{,}\frac{1}{2}\left(
-1+\mathbf{x}_{g}^{2}\right)  ,\frac{1}{2}\left(  1+\mathbf{x}_{g}^{2}\right)
\right)  .\tag{2.11}%
\end{equation}
Below we give the expressions of $\omega_{g}\left(  \mathbf{x}\right)  $ and
$\mathbf{x}_{g}$ for subgroups in (\ref{ga}) :

1. Translation subgroup $N\ni g=n\left(  \mathbf{b}\right)  $%
\begin{equation}
\omega_{g}\left(  \mathbf{x}\right)  =1,\quad\mathbf{x}_{g}=\mathbf{x}%
-\mathbf{b.} \tag{2.12}%
\end{equation}

2. Dilatation subgroup. $A\ni g=a\left(  \lambda\right)  $%
\begin{equation}
\omega_{g}\left(  \mathbf{x}\right)  =\lambda,\quad\mathbf{x}_{g}%
=\frac{\mathbf{x}}{\lambda}. \tag{2.13}%
\end{equation}

3. Rotation subgroup. $M\ni g=m$%
\begin{equation}
\omega_{g}\left(  \mathbf{x}\right)  =1,\quad\mathbf{x}_{g}=m^{-1}\mathbf{x.}
\tag{2.14}%
\end{equation}

4. The subgroup of special conformal transformations $\overline{N}\ni
g=\overline{n}\left(  \mathbf{c}\right)  $%
\begin{equation}
\omega_{g}\left(  \mathbf{x}\right)  =1+\mathbf{c}^{2}\mathbf{x}%
^{2}-2\mathbf{cx},\quad\mathbf{x}_{g}=\frac{\mathbf{x-cx}^{2}}{1+\mathbf{c}%
^{2}\mathbf{x}^{2}-2\mathbf{cx}}. \tag{2.15}%
\end{equation}

The representation $T^{\sigma}\left(  g\right)  $ of $SO_{0}\left(
d+1,1\right)  $ is irreducible if $\sigma$ is a non-integer or $\sigma$ is an
integer from the interval $-d-1\leq\sigma<0$. The irreducible representations
$T^{\sigma}\left(  g\right)  $ can be extended to unitary ( irreducible )
representations of $SO_{0}\left(  d+1,1\right)  $ for the following values of
$\sigma$ :

1. If $\sigma=-\frac{d}{2}+i\rho,$\thinspace$\rho\in R$ (principal series).

2. If $-d<\sigma<0$ (complementary series).

3. If $\sigma=-d-p$ with $p\in Z_{+}$ (discrete series).

We restrict the discussion to the principal series of class 1 unitary
irreducible representations of $SO_{0}\left(  d+1,1\right)  $, that are
relevant to our purpose. The principal series representations $T^{\sigma
}\left(  g\right)  ,\sigma=-\frac{d}{2}+i\rho$ act on the space $L^{2}(R^{d})
$ of square-integrable functions on $R^{d}$. The representations $T^{\sigma}
$\ and $T^{\widetilde{\sigma}}$,\ where $\widetilde{\sigma}=-d-\sigma$, are equivalent.

The principal series representation $T^{\sigma}$ can also be realized in the
space $L^{2}(R^{d})$ of square-integrable functions on the $d$-dimensional
momentum space $R^{d}$. This realization is obtained by passing from functions
$f\left(  \mathbf{x}\right)  $ to their Fourier transform
\begin{equation}
\varphi\left(  \mathbf{p}\right)  =\int_{R^{d}}e^{i\mathbf{px}}f\left(
\mathbf{x}\right)  d\mathbf{x.} \tag{2.16}%
\end{equation}
We find that%
\begin{equation}
\left(  T^{\sigma}\left(  g\right)  \varphi\right)  \left(  \mathbf{p}\right)
=\int_{R^{d}}K^{\sigma}\left(  \mathbf{p,p}^{^{\prime}},g\right)
\varphi\left(  \mathbf{p}^{^{\prime}}\right)  d\mathbf{p}^{^{\prime}},
\tag{2.17}%
\end{equation}
where%
\begin{equation}
K^{\sigma}\left(  \mathbf{p,p}^{^{\prime}},g\right)  =\frac{1}{\left(
2\pi\right)  ^{d}}\int_{R^{d}}e^{i\mathbf{px-}i\mathbf{p}^{^{\prime}%
}\mathbf{x}_{g}}\omega_{g}\left(  \mathbf{x}\right)  ^{\sigma}d\mathbf{x.}
\tag{2.18}%
\end{equation}
Restricting $T^{\sigma}\left(  g\right)  $ to the subgroup $ISO(d)$, we obtain%
\begin{align}
\left(  T^{\sigma}\left(  n\left(  \mathbf{b}\right)  \right)  \varphi\right)
\left(  \mathbf{p}\right)   &  =e^{i\mathbf{pb}}\varphi\left(  \mathbf{p}%
\right)  ,\nonumber\\
\left(  T^{\sigma}\left(  m\right)  \varphi\right)  \left(  \mathbf{p}\right)
&  =\varphi\left(  m^{-1}\mathbf{p}\right)  ,\nonumber\\
&  \tag{2.19}%
\end{align}
where $n\left(  \mathbf{b}\right)  \in N$ and $m\in M$. Thus, the restriction
of the representation $T^{\sigma}\left(  g\right)  $ of the group
$SO_{0}\left(  d+1,1\right)  $ to the subgroup $ISO(d)$ is a reducible
representation of this subgroup.

\section{Euclidean field theory with conformal symmetry}

We begin this section by defining the Euclidean Fock space $\mathcal{E}$ for
scalar bosons \cite{fel73}, \cite{ost73}. The Euclidean one-particle space
$\mathcal{E}_{1}$ is represented as $L^{2}(R^{d})$ and the Euclidean Fock
space $\mathcal{E}$ is%
\begin{equation}
\mathcal{E=\oplus}_{n=0}^{\infty}\;\mathcal{E}_{n},\tag{3.1}%
\end{equation}
where $\mathcal{E}_{0}=C$ \ and $\mathcal{E}_{n}$ is the $n$-fold symmetric
tensor product of $\mathcal{E}_{1}$%
\begin{equation}
\mathcal{E}_{n}=\left(  \mathcal{E}_{1}\otimes\cdots\otimes\mathcal{E}%
_{1}\right)  _{sym}.\tag{3.2}%
\end{equation}
A unitary representation of the Euclidean group $ISO(d)$ is defined on
$\mathcal{E}$ by%
\begin{equation}
\left(  U\left(  m,\mathbf{b}\right)  \varphi\right)  _{n}\left(
\mathbf{p}_{1},\ldots,\mathbf{p}_{n}\right)  =e^{i\left(  \mathbf{p}%
_{1}+\ldots+\mathbf{p}_{n}\right)  \mathbf{b}}\varphi_{n}\left(
m^{-1}\mathbf{p}_{1},\ldots,m^{-1}\mathbf{p}_{n}\right)  ,\tag{3.3}%
\end{equation}
where $\left(  m.\mathbf{b}\right)  \in\,$ $ISO(d)$ with $m\in SO(d)$ and
$\mathbf{b\in\,}R^{d}$. It should be noted that the representation of the
group $ISO(d)$ defined on $\mathcal{E}_{1}$ is nothing but the unitary
reducible representation (2.19).

In Euclidean field theory with dynamic symmetry $SO_{0}\left(  d+1,1\right)
$\ we assume that the representations of $SO_{0}\left(  d+1,1\right)  $ acting
on $\mathcal{E}^{in}$ and $\mathcal{E}^{out}$ are induced by principal series
representations $T^{\sigma}\left(  g\right)  $ and $T^{\widetilde{\sigma}%
}\left(  g\right)  ,\,\widetilde{\sigma}=-d-\sigma$, respectively, i.e.%
\begin{align}
&  \left(  T^{\sigma_{1}\ldots\sigma_{n}}\left(  g\right)  \varphi\right)
_{n}\left(  \mathbf{p}_{1},\ldots,\mathbf{p}_{n}\right) \nonumber\\
&  =\int_{R^{d}}\cdots\int_{R^{d}}K^{\sigma_{1}}\left(  \mathbf{p}%
_{1}\mathbf{,p}_{1}^{^{\prime}},g\right)  \cdots\,K^{\sigma_{n}}\left(
\mathbf{p}_{n}\mathbf{,p}_{n}^{^{\prime}},g\right)  \varphi_{n}\left(
\mathbf{p}_{1}^{\prime},\ldots,\mathbf{p}_{n}^{\prime}\right)  d\mathbf{p}%
_{1}^{^{\prime}}\cdots\,d\mathbf{p}_{n}^{^{\prime}}\nonumber\\
&  \tag{3.4}%
\end{align}
and%
\begin{align}
&  \left(  T^{\widetilde{\sigma}_{1}\ldots\widetilde{\sigma}_{n}}\left(
g\right)  \varphi\right)  _{n}\left(  \mathbf{p}_{1},\ldots,\mathbf{p}%
_{n}\right) \nonumber\\
&  =\int_{R^{d}}\cdots\int_{R^{d}}K^{\widetilde{\sigma}_{1}}\left(
\mathbf{p}_{1}\mathbf{,p}_{1}^{^{\prime}},g\right)  \cdots\,K^{\widetilde
{\sigma}_{n}}\left(  \mathbf{p}_{n}\mathbf{,p}_{n}^{^{\prime}},g\right)
\varphi_{n}\left(  \mathbf{p}_{1}^{\prime},\ldots,\mathbf{p}_{n}^{\prime
}\right)  d\mathbf{p}_{1}^{^{\prime}}\cdots\,d\mathbf{p}_{n}^{^{\prime}%
}.\nonumber\\
&  \tag{3.5}%
\end{align}
In other words, $T^{\sigma_{1}\ldots\sigma_{n}}=$\ $T^{\sigma_{1}}%
\otimes\cdots\otimes T^{\sigma_{n}}$\ and $T^{\widetilde{\sigma}_{1}%
\ldots\widetilde{\sigma}_{n}}=T^{\widetilde{\sigma}_{1}}\otimes\cdots\otimes
T^{\widetilde{\sigma}_{n}}$.\ Since $L^{2}(R^{d})$ is the carrier space of the
principal series representations, the representations of the group
$SO_{0}\left(  d+1,1\right)  $ defined on $\mathcal{E}^{in}$ and
$\mathcal{E}^{out}$ are unitary. Moreover, the equality $\mathcal{E=E}%
^{in}=\mathcal{E}^{out}$ holds.

The aim of this section is to study the intertwining operator between the
representations $T^{\sigma_{1}\ldots\sigma_{n_{1}}}\left(  g\right)  $ and
$T^{\widetilde{\sigma}_{n_{1}+1}\ldots\widetilde{\sigma}_{n}}\left(  g\right)
$, where $n=n_{1}+n_{2}$. In momentum space realization, the operators of
these representations are integral operators with kernel. Therefore, it is
more convenient to work with the position space realization of these
representations instead of dealing with the momentum space realization.

Let $\mathcal{S}$ be an intertwining operator between the representations%
\begin{align}
&  \left(  T^{\sigma_{1}\ldots\sigma_{n_{1}}}\left(  g\right)  f\right)
_{n_{1}}\left(  \mathbf{x}_{1},\ldots,\mathbf{x}_{n_{1}}\right) \nonumber\\
&  =\omega_{g}\left(  \mathbf{x}_{1}\right)  ^{\sigma_{1}}\cdots\,\omega
_{g}\left(  \mathbf{x}_{n_{1}}\right)  ^{\sigma_{n_{1}}}f_{n_{1}}\left(
\mathbf{x}_{1g},\ldots,\mathbf{x}_{n_{1}g}\right)  \tag{3.6}%
\end{align}
and%
\begin{align}
&  \left(  T^{\widetilde{\sigma}_{n_{1}+1}\ldots\widetilde{\sigma}_{n}}\left(
g\right)  f\right)  _{n_{2}}\left(  \mathbf{x}_{n_{1}+1},\ldots,\mathbf{x}%
_{n}\right) \nonumber\\
&  =\omega_{g}\left(  \mathbf{x}_{n_{1}+1}\right)  ^{\widetilde{\sigma}%
_{n_{1}+1}}\cdots\,\omega_{g}\left(  \mathbf{x}_{n}\right)  ^{\widetilde
{\sigma}_{n}}f_{n_{2}}\left(  \mathbf{x}_{n_{1}+1g},\ldots,\mathbf{x}%
_{ng}\right)  , \tag{3.7}%
\end{align}
which means that%
\begin{equation}
\mathcal{S}T^{\widetilde{\sigma}_{n_{1}+1}\ldots\widetilde{\sigma}_{n}}\left(
g\right)  =T^{\sigma_{1}\ldots\sigma_{n_{1}}}\left(  g\right)  \mathcal{S}%
\text{\thinspace},\text{\quad}\forall g\in SO_{0}\left(  d+1,1\right)  .
\tag{3.8}\label{int1}%
\end{equation}
Such an operator is defined by a kernel%
\begin{align}
&  \left(  \mathcal{S}f\right)  _{n_{1}}\left(  \mathbf{x}_{1},\ldots
,\mathbf{x}_{n_{1}}\right) \nonumber\\
&  =\int_{R^{d}}\cdots\int_{R^{d}}S_{n}\left(  \mathbf{x}_{1},\ldots
,\mathbf{x}_{n}\right)  f_{n_{2}}\left(  \mathbf{x}_{n_{1}+1},\ldots
,\mathbf{x}_{n}\right)  d\mathbf{x}_{n_{1}+1}\cdots\,d\mathbf{x}_{n}\,.
\tag{3.9}%
\end{align}
The intertwining relation (\ref{int1}) implies%
\begin{equation}
\left(  \mathcal{S}T^{\widetilde{\sigma}_{n_{1}+1}\ldots\widetilde{\sigma}%
_{n}}\left(  g\right)  f\right)  _{n_{1}}\left(  \mathbf{x}_{1},\ldots
,\mathbf{x}_{n_{1}}\right)  =\left(  T^{\sigma_{1}\ldots\sigma_{n_{1}}}\left(
g\right)  \mathcal{S}f\right)  _{n_{1}}\left(  \mathbf{x}_{1},\ldots
,\mathbf{x}_{n_{1}}\right)  . \tag{3.10}%
\end{equation}
It follows from here that%
\begin{align}
&  \int_{R^{d}}\cdots\int_{R^{d}}S_{n}\left(  \mathbf{x}_{1},\ldots
,\mathbf{x}_{n}\right)  \omega_{g}\left(  \mathbf{x}_{n_{1}+1}\right)
^{\widetilde{\sigma}_{n_{1}+1}}\cdots\,\omega_{g}\left(  \mathbf{x}%
_{n}\right)  ^{\widetilde{\sigma}_{n}}\nonumber\\
&  \times f_{n_{2}}\left(  \mathbf{x}_{n_{1}+1g},\ldots,\mathbf{x}%
_{ng}\right)  d\mathbf{x}_{n_{1}+1}\cdots\,d\mathbf{x}_{n}\,\nonumber\\
&  =\omega_{g}\left(  \mathbf{x}_{1}\right)  ^{\sigma_{1}}\cdots\,\omega
_{g}\left(  \mathbf{x}_{n_{1}}\right)  ^{\sigma_{n_{1}}}\int_{R^{d}}\cdots
\int_{R^{d}}S_{n}\left(  \mathbf{x}_{1g},\ldots,\mathbf{x}_{n_{1}g}%
,\mathbf{x}_{n_{1}+1}^{\prime},\ldots,\mathbf{x}_{n}^{\prime}\right)
\nonumber\\
&  f_{n_{2}}\left(  \mathbf{x}_{n_{1}+1}^{\prime},\ldots,\mathbf{x}%
_{n}^{\prime}\right)  d\mathbf{x}_{n_{1}+1}^{\prime}\cdots\,d\mathbf{x}%
_{n}^{\prime}. \tag{3.11}%
\end{align}
Performing the change of variables
\begin{equation}
\mathbf{x}_{i}^{\prime}=\mathbf{x}_{ig}\,,\quad d\mathbf{x}_{i}^{\prime
}=\omega_{g}\left(  \mathbf{x}_{i}\right)  ^{-d}d\mathbf{x}_{i}\,,\quad
i=n_{1}+1,\ldots,n, \tag{3.12}%
\end{equation}
on the right side of (3.11) and taking into account the arbitrariness of the
function\thinspace\ $f_{n_{2}}$ we find the following constraint on the Green
functions $S_{n}$:%
\begin{equation}
S_{n}\left(  \mathbf{x}_{1},\ldots,\mathbf{x}_{n}\right)  =\omega_{g}\left(
\mathbf{x}_{1}\right)  ^{\sigma_{1}}\cdots\,\omega_{g}\left(  \mathbf{x}%
_{n}\right)  ^{\sigma_{n}}S_{n}\left(  \mathbf{x}_{1g},\ldots,\mathbf{x}%
_{ng}\right)  . \tag{3.13}%
\end{equation}

It is well known \cite{pol70} that conformal symmetry allows one to determine
the two- and three-point functions up to a constant factor. Below we provide a
step-by-step calculation of these Green functions.

Let's start with the two-point function. In this case,%
\begin{equation}
S_{2}\left(  \mathbf{x}_{1},\mathbf{x}_{2}\right)  =\omega_{g}\left(
\mathbf{x}_{1}\right)  ^{\sigma_{1}}\omega_{g}\left(  \mathbf{x}_{2}\right)
^{\sigma_{2}}S_{2}\left(  \mathbf{x}_{1g},\mathbf{x}_{2g}\right)
.\tag{3.14}\label{cov2}%
\end{equation}
By choosing $g=n\left(  \mathbf{b}\right)  $ and letting $\mathbf{b}%
=\mathbf{x}_{2}$, we see that $S_{2}\left(  \mathbf{x}_{1},\mathbf{x}%
_{2}\right)  $ is a function of $\mathbf{x}_{1}-\mathbf{x}_{2}$ only
\begin{equation}
S_{2}\left(  \mathbf{x}_{1},\mathbf{x}_{2}\right)  =S_{2}\left(
\mathbf{x}_{1}-\mathbf{x}_{2}\right)  ,\tag{3.15}\label{tr2}%
\end{equation}
where $S_{2}\left(  \mathbf{x}\right)  =S_{2}\left(  \mathbf{x},\mathbf{0}%
\right)  $. Inserting (\ref{tr2}) into (\ref{cov2}), we obtain%
\begin{equation}
S_{2}\left(  \mathbf{x}_{1}-\mathbf{x}_{2}\right)  =\omega_{g}\left(
\mathbf{x}_{1}\right)  ^{\sigma_{1}}\omega_{g}\left(  \mathbf{x}_{2}\right)
^{\sigma_{2}}S_{2}\left(  \mathbf{x}_{1g}-\mathbf{x}_{2g}\right)
.\tag{3.16}\label{cov2a}%
\end{equation}
Let now $g=a\left(  \lambda\right)  $. By putting $\lambda=\left\Vert
\mathbf{x}_{1}-\mathbf{x}_{2}\right\Vert $ we find
\begin{equation}
S_{2}\left(  \mathbf{x}\right)  =\left\Vert \mathbf{x}\right\Vert ^{\sigma
_{1}+\sigma_{2}}S_{2}\left(  \frac{\mathbf{x}}{\left\Vert \mathbf{x}%
\right\Vert }\right)  ,\tag{3.17}\label{gr2a}%
\end{equation}
where $\left\Vert \mathbf{x}\right\Vert =\sqrt{x_{1}^{2}+x_{2}^{2}%
+\cdots+x_{n}^{2}}$. If we insert (\ref{gr2a}) into (\ref{cov2a}) and put
$g=m$, we have%
\begin{equation}
S_{2}\left(  \mathbf{x}\right)  =\left\Vert \mathbf{x}\right\Vert ^{\sigma
_{1}+\sigma_{2}}S_{2}\left(  \frac{m^{-1}\mathbf{x}}{\left\Vert \mathbf{x}%
\right\Vert }\right)  .\tag{3.18}%
\end{equation}
Let us recall that every vector $\mathbf{x}\in R^{d}$ is represented as
$\mathbf{x}=\left\Vert \mathbf{x}\right\Vert m_{\mathbf{x}}\mathbf{e}$, where
$m_{\mathbf{x}}\in SO\left(  d\right)  $ and $\mathbf{e=}\left(
0,\ldots,0,1\right)  $. The matrix $m_{\mathbf{x}}$ is independent of the
length of $\mathbf{x}$, i.e. it depends only on $\mathbf{x}\,/\left\Vert
\mathbf{x}\right\Vert $. Choosing $m=m_{\mathbf{x}}$, we have%
\begin{equation}
S_{2}\left(  \mathbf{x}\right)  =S_{2}\left(  \mathbf{e}\right)  \left\Vert
\mathbf{x}\right\Vert ^{\sigma_{1}+\sigma_{2}}.\tag{3.19}\label{gr2b}%
\end{equation}
All that remains is to use special conformal transformations. Inserting
(\ref{gr2b}) in (\ref{cov2a}) and choosing $g=\overline{n}\left(
\mathbf{c}\right)  $, we obtain%
\begin{equation}
\left\Vert \mathbf{x}_{1}-\mathbf{x}_{2}\right\Vert ^{\sigma_{1}+\sigma_{2}%
}=\left(  \frac{1+\mathbf{c}^{2}\mathbf{x}_{1}^{2}-2\mathbf{cx}_{1}%
}{1+\mathbf{c}^{2}\mathbf{x}_{2}^{2}-2\mathbf{cx}_{2}}\right)  ^{\frac
{\sigma_{1}-\sigma_{2}}{2}}\left\Vert \mathbf{x}_{1}-\mathbf{x}_{2}\right\Vert
^{\sigma_{1}+\sigma_{2}}.\tag{3.20}%
\end{equation}
This is only possible if $\sigma_{1}=\sigma_{2}$. Hence%
\begin{equation}
S_{2}\left(  \mathbf{x}_{1},\mathbf{x}_{2}\right)  =C\left\Vert \mathbf{x}%
_{1}-\mathbf{x}_{2}\right\Vert ^{2\sigma}\text{\qquad if\quad\ }\sigma
_{1}=\sigma_{2}=\sigma,\tag{3.21}%
\end{equation}
where $C$ is a constant coefficient, which is not determined by conformal
symmetry.\qquad\qquad

A similar analysis can be done on three-point functions. In this case,
\begin{equation}
S_{3}\left(  \mathbf{x}_{1},\mathbf{x}_{2},\mathbf{x}_{3}\right)  =\omega
_{g}\left(  \mathbf{x}_{1}\right)  ^{\sigma_{1}}\omega_{g}\left(
\mathbf{x}_{2}\right)  ^{\sigma_{2}}\omega_{g}\left(  \mathbf{x}_{3}\right)
^{\sigma_{3}}S_{3}\left(  \mathbf{x}_{1g},\mathbf{x}_{2g},\mathbf{x}%
_{3g}\right)  . \tag{3.22}\label{cov3}%
\end{equation}
Translation covariance imposes the following constraint on $S_{3}$%
\begin{equation}
S_{3}\left(  \mathbf{x}_{1},\mathbf{x}_{2},\mathbf{x}_{3}\right)
=S_{3}\left(  \mathbf{x}_{1}-\mathbf{x}_{3},\mathbf{x}_{2}-\mathbf{x}%
_{3}\right)  , \tag{3.23}\label{tr3}%
\end{equation}
where $S_{3}\left(  \mathbf{x}_{1},\mathbf{x}_{2}\right)  =S_{3}\left(
\mathbf{x}_{1},\mathbf{x}_{2},\mathbf{0}\right)  $. Inserting (\ref{tr3}) into
(\ref{cov3}), we obtain%
\begin{align}
&  S_{3}\left(  \mathbf{x}_{1}-\mathbf{x}_{3},\mathbf{x}_{2}-\mathbf{x}%
_{3}\right) \nonumber\\
&  =\omega_{g}\left(  \mathbf{x}_{1}\right)  ^{\sigma_{1}}\omega_{g}\left(
\mathbf{x}_{2}\right)  ^{\sigma_{2}}\omega_{g}\left(  \mathbf{x}_{3}\right)
^{\sigma_{3}}S_{3}\left(  \mathbf{x}_{1g}-\mathbf{x}_{3g},\mathbf{x}%
_{2g}-\mathbf{x}_{3g}\right)  . \tag{3.24}%
\end{align}
Scale and rotation covariance implies that%
\begin{equation}
S_{3}\left(  \mathbf{x}_{1},\mathbf{x}_{2}\right)  =\left\Vert \mathbf{x}%
_{2}\right\Vert ^{\sigma_{1}+\sigma_{2}+\sigma_{3}}S_{3}\left(  \frac
{m_{\mathbf{x}_{2}}^{-1}\mathbf{x}_{1}}{\left\Vert \mathbf{x}_{2}\right\Vert
}\right)  , \tag{3.25}\label{gr3a}%
\end{equation}
where $S_{3}\left(  \mathbf{x}\right)  =S_{3}\left(  \mathbf{x},\mathbf{e}%
\right)  $. We now insert (\ref{gr3a}) in (3.24) and make use of covariance
under special conformal transformations $g=\overline{n}\left(  \mathbf{c}%
\right)  $. By putting
\begin{equation}
\mathbf{x}_{3}=\mathbf{0,\quad x}_{2}=\mathbf{e,\quad c}=\frac{2\mathbf{x}%
_{1}}{\mathbf{x}_{1}^{2}}-\mathbf{e,} \tag{3.26}%
\end{equation}
we obtain%
\begin{equation}
S_{3}\left(  \mathbf{x}\right)  =2^{-\sigma_{1}+\sigma_{2}-\sigma_{3}}%
S_{3}\left(  2\mathbf{e}\right)  \left\Vert \mathbf{x}\right\Vert ^{\sigma
_{1}-\sigma_{2}+\sigma_{3}}\left\Vert \mathbf{e-x}\right\Vert ^{\sigma
_{1}+\sigma_{2}-\sigma_{3}} \tag{3.27}\label{gr3b}%
\end{equation}
( with $\mathbf{x}_{1}$ replaced by $\mathbf{x}$ ). Then according to
(\ref{gr3b}), (\ref{gr3a}) and (\ref{tr3}) we have%
\begin{equation}
S_{3}\left(  \mathbf{x}_{1},\mathbf{x}_{2},\mathbf{x}_{3}\right)
=C_{3}\left\Vert \mathbf{x}_{1}-\mathbf{x}_{3}\right\Vert ^{\sigma_{1}%
-\sigma_{2}+\sigma_{3}}\left\Vert \mathbf{x}_{2}-\mathbf{x}_{3}\right\Vert
^{-\sigma_{1}+\sigma_{2}+\sigma_{3}}\left\Vert \mathbf{x}_{1}-\mathbf{x}%
_{2}\right\Vert ^{\sigma_{1}+\sigma_{2}-\sigma_{3}}, \tag{3.28}%
\end{equation}
where $C_{3}$ is a three-point structure constant depending on $\sigma
_{1},\sigma_{2},\sigma_{3}$.

It has long been known that three-point functions of conformal field theory
are related to the Clebsch-Gordan kernels of the conformal group (see
\cite{dob77} and references therein). This fact is more transparent in the
definition (\ref{int1}). Moreover, by decomposing the tensor product of two
principal series representations into irreducibles, $n$-point functions can be
written in terms of Clebsch-Gordan kernels. This will be given explicitly for
four-point functions in two dimensions in the next section.

\section{Four-point functions in two dimensions}

To outline our technique, we restrict ourselves, for the sake of simplicity,
to the four point functions in two dimensions. Instead of working with the
group $SO_{0}\left(  3,1\right)  $, we use its universal covering group
$SL(2,C)$, consisting of all complex unimodular $2\times2$ matrices. It is
known that any representation of $SL(2,C)$ is a single- or double- valued
representation of $SO_{0}\left(  3,1\right)  $. We denote an element of
$SL(2,C)$ by%
\begin{equation}
g=\left(
\begin{array}
[c]{cc}%
g_{11} & g_{12}\\
g_{21} & g_{22}%
\end{array}
\right)  ,\qquad g_{11}g_{22}-g_{12}g_{21}=1. \tag{4.1}%
\end{equation}
The principal series representations of $SL(2,C)$ \cite{naim2} are
characterized by the pairs of numbers $\chi=$ $\left(  s,\sigma\right)  $
where $s$ is an integer or half- integer, and $\sigma=-1+i\rho,\quad\rho\in
R$. These representations can be realized in a Hilbert space $L^{2}(C)$%
\begin{equation}
\left(  T^{\chi}\left(  g\right)  f\right)  \left(  z\right)  =\left\vert
g_{12}z+g_{22}\right\vert ^{2\sigma-2s}\left(  g_{12}z+g_{22}\right)
^{2s}f\left(  \frac{g_{11}z+g_{21}}{g_{12}z+g_{22}}\right)  . \tag{4.2}%
\end{equation}
The representations $\chi=$ $\left(  s,\sigma\right)  $ and $\widetilde{\chi
}=$ $\left(  -s,\widetilde{\sigma}\right)  $, $\widetilde{\sigma}=-2-\sigma$,
are equivalent. We will restrict ourselves to scalar principal series
representations $\chi=$ $\left(  0,\sigma\right)  $ and will denote the
corresponding operators by $T^{\sigma}$. A brief overview of Naimark's results
on the decomposition of tensor products of scalar principal series
representations is given in the Appendix.

We shall construct intertwining operators between representations%
\begin{align}
&  \left(  T^{\sigma_{1}\sigma_{2}}\left(  g\right)  f\right)  \left(
z_{1},z_{2}\right)  \nonumber\\
&  =\left\vert g_{12}z_{1}+g_{22}\right\vert ^{2\sigma_{1}}\left\vert
g_{12}z_{2}+g_{22}\right\vert ^{2\sigma_{2}}f\left(  \frac{g_{11}z_{1}+g_{21}%
}{g_{12}z_{1}+g_{22}},\frac{g_{11}z_{2}+g_{21}}{g_{12}z_{2}+g_{22}}\right)
\tag{4.3}%
\end{align}
and%
\begin{align}
&  \left(  T^{\widetilde{\sigma}_{3}\widetilde{\sigma}_{4}}\left(  g\right)
f\right)  \left(  z_{3},z_{4}\right)  \nonumber\\
&  =\left\vert g_{12}z_{3}+g_{22}\right\vert ^{2\widetilde{\sigma}_{3}%
}\left\vert g_{12}z_{4}+g_{22}\right\vert ^{2\widetilde{\sigma}_{4}}f\left(
\frac{g_{11}z_{3}+g_{21}}{g_{12}z_{3}+g_{22}},\frac{g_{11}z_{4}+g_{21}}%
{g_{12}z_{4}+g_{22}}\right)  .\tag{4.4}%
\end{align}
By decomposing $T^{\sigma_{1}\sigma_{2}}\left(  g\right)  $ into irreducibles
(see Appendix), the problem reduces to the much simpler problem of finding an
intertwining operator between the representations $T^{\widetilde{\sigma}%
_{3}\widetilde{\sigma}_{4}}$ and $T^{\chi}$. Thus, we get that the operator
$\mathcal{S}$ with the intertwining property%
\begin{equation}
\mathcal{S}T^{\widetilde{\sigma}_{3}\widetilde{\sigma}_{4}}\left(  g\right)
=T^{\sigma_{1}\sigma_{2}}\left(  g\right)  \mathcal{S},\tag{4.5}%
\end{equation}
can be written as an integral operator%
\begin{equation}
\left(  \mathcal{S}f\right)  \left(  z_{1},z_{2}\right)  =\int_{C^{2}}%
S_{4}\left(  z_{1},z_{2},z_{3},z_{4}\right)  f\left(  z_{3},z_{4}\right)
dz_{3}dz_{4},\tag{4.6}%
\end{equation}
with kernel%
\begin{align}
&  S_{4}\left(  z_{1},z_{2},z_{3},z_{4}\right)  \nonumber\\
&  =\frac{1}{4\pi^{4}}%
%TCIMACRO{\tsum \limits_{s=-\infty}^{\infty}}%
%BeginExpansion
{\textstyle\sum\limits_{s=-\infty}^{\infty}}
%EndExpansion%
%TCIMACRO{\dint \limits_{-\infty}^{\infty}}%
%BeginExpansion
{\displaystyle\int\limits_{-\infty}^{\infty}}
%EndExpansion
d\rho\left(  s^{2}+\rho^{2}\right)  \overline{C(\sigma_{1},\sigma_{2},\chi
)}C(\widetilde{\sigma}_{3},\widetilde{\sigma}_{4},\chi)\nonumber\\
&  \times S_{4}^{\chi}\left(  z_{1},z_{2},z_{3},z_{4}\right)  \tag{4.7}%
\end{align}
(the bar denotes complex conjugate). Here $C$ is the three-point structure
constant, and $S_{4}^{\chi}$ is an $s$-channel four-point conformal block%
\begin{equation}
S_{4}^{\chi}\left(  z_{1},z_{2},z_{3},z_{4}\right)  =\int_{C}\overline
{N\left(  z_{1}\sigma_{1}z_{2}\sigma_{2};z\chi\right)  }N\left(
z_{3}\widetilde{\sigma}_{3}z_{4}\widetilde{\sigma}_{4};z\chi\right)
dz.\tag{4.8}\label{cb}%
\end{equation}
Inserting for the kernels $N$ in (\ref{cb}) its expression (A.3), we have%
\begin{align}
S_{4}^{\chi}\left(  z_{1},z_{2},z_{3},z_{4}\right)   &  =\left\vert
z_{3}-z_{4}\right\vert ^{s-\sigma+\sigma_{3}+\sigma_{4}}\left(  z_{3}%
-z_{4}\right)  ^{-s}\nonumber\\
&  \times\left\vert z_{1}-z_{2}\right\vert ^{s+\sigma+\sigma_{1}+\sigma_{2}%
+2}\overline{\left(  z_{1}-z_{2}\right)  }^{-s}J,\tag{4.9}%
\end{align}
where%
\begin{equation}
J=\int_{C}dz%
%TCIMACRO{\tprod \limits_{i=1}^{4}}%
%BeginExpansion
{\textstyle\prod\limits_{i=1}^{4}}
%EndExpansion
\left(  z-z_{i}\right)  ^{-\alpha_{i}}\overline{\left(  z-z_{i}\right)
}^{-\nu_{i}},\tag{4.10}\label{integ}%
\end{equation}
with%
\begin{align}
2\alpha_{1} &  =s+\sigma-\sigma_{1}+\sigma_{2}+2,\qquad2\nu_{1}=-s+\sigma
-\sigma_{1}+\sigma_{2}+2,\nonumber\\
2\alpha_{2} &  =s+\sigma+\sigma_{1}-\sigma_{2}+2,\qquad2\nu_{2}=-s+\sigma
+\sigma_{1}-\sigma_{2}+2,\nonumber\\
2\alpha_{3} &  =-s-\sigma-\sigma_{3}+\sigma_{4},\qquad\quad2\nu_{3}%
=s-\sigma-\sigma_{3}+\sigma_{4},\nonumber\\
2\alpha_{4} &  =-s-\sigma+\sigma_{3}-\sigma_{4},\qquad\quad2\nu_{4}%
=s-\sigma+\sigma_{3}-\sigma_{4}.\nonumber\\
& \tag{4.11}%
\end{align}
These parameters are subject to the condition%
\begin{equation}%
%TCIMACRO{\tsum \limits_{i=1}^{4}}%
%BeginExpansion
{\textstyle\sum\limits_{i=1}^{4}}
%EndExpansion
\alpha_{i}=%
%TCIMACRO{\tsum \limits_{i=1}^{4}}%
%BeginExpansion
{\textstyle\sum\limits_{i=1}^{4}}
%EndExpansion
\nu_{i}=2,\quad\alpha_{i}-\nu_{i}\in Z.\tag{4.12}%
\end{equation}
From (4.9) we obtain%
\begin{align}
S_{4}^{\chi}\left(  z_{1},z_{2},z_{3},z_{4}\right)   &  =\left\vert
g_{12}z_{1}+g_{22}\right\vert ^{2\sigma_{1}}\cdots\left\vert g_{12}%
z_{4}+g_{22}\right\vert ^{2\sigma_{4}}\nonumber\\
&  \times S_{4}^{\chi}\left(  \frac{g_{11}z_{1}+g_{21}}{g_{12}z_{1}+g_{22}%
},\ldots,\frac{g_{11}z_{4}+g_{21}}{g_{12}z_{4}+g_{22}}\right)  ,\tag{4.13}%
\end{align}
i.e. conformal blocks $S_{4}^{\chi}$ are subject to the same conformal
symmetry constraints as the four-point functions.

The integral (\ref{integ}) was evaluated in \cite{dol11}, and hence we have%
\begin{align}
&  S_{4}^{\chi}\left(  z_{1},z_{2},z_{3},z_{4}\right) \nonumber\\
&  =\pi z_{12}^{\alpha_{5}}z_{23}^{\alpha_{1}+\alpha_{4}-1}z_{31}^{\alpha
_{2}-1}z_{24}^{-\alpha_{4}}z_{34}^{\alpha_{6}}\overline{z}_{12}^{\nu_{5}%
}\overline{z}_{23}^{\nu_{1}+\nu_{4}-1}\overline{z}_{31}^{\nu_{2}-1}%
\overline{z}_{24}^{-\nu_{4}}\overline{z}_{34}^{\nu_{6}}\Phi\left(
\eta,\overline{\eta}\right)  ,\nonumber\\
&  \tag{4.14}%
\end{align}
where $z_{ij}=z_{i}-z_{j}$ and\
\begin{align}
2\alpha_{5}  &  =-s-\sigma+\sigma_{1}+\sigma_{2},\qquad\ 2\nu_{5}%
=s-\sigma+\sigma_{1}+\sigma_{2},\nonumber\\
2\alpha_{6}  &  =-s-\sigma+\sigma_{3}+\sigma_{4},\qquad\ 2\nu_{6}%
=s-\sigma+\sigma_{3}+\sigma_{4}.\nonumber\\
&  \tag{4.15}%
\end{align}
The function $\Phi$ is expressed in terms of products of ordinary
hypergeometric functions%
\begin{align}
\Phi\left(  \eta,\overline{\eta}\right)   &  =AF\left(  1-\alpha_{2}%
,\alpha_{4};\alpha_{3}+\alpha_{4};\eta\right)  F\left(  1-\nu_{2},\nu_{4}%
;\nu_{3}+\nu_{4};\overline{\eta}\right) \nonumber\\
&  +B\eta^{\alpha_{1}+\alpha_{2}-1}\overline{\eta}^{\nu_{1}+\nu_{2}-1}F\left(
1-\alpha_{3},\alpha_{1};\alpha_{1}+\alpha_{2};\eta\right) \nonumber\\
&  \times F\left(  1-\nu_{3},\nu_{1};\nu_{1}+\nu_{2};\overline{\eta}\right)  ,
\tag{4.16}%
\end{align}
with%
\begin{equation}
\eta=\frac{z_{12}z_{34}}{z_{13}z_{24}}, \tag{4.18}%
\end{equation}
and%
\begin{align}
A  &  =\frac{\Gamma\left(  1-\alpha_{1}\right)  \Gamma\left(  1-\alpha
_{2}\right)  \Gamma\left(  \alpha_{1}+\alpha_{2}-1\right)  }{\Gamma\left(
\nu_{1}\right)  \Gamma\left(  \nu_{2}\right)  \Gamma\left(  2-\nu_{1}-\nu
_{2}\right)  },\tag{4.19}\\
B  &  =\frac{\Gamma\left(  1-\alpha_{3}\right)  \Gamma\left(  1-\alpha
_{4}\right)  \Gamma\left(  \alpha_{3}+\alpha_{4}-1\right)  }{\Gamma\left(
\nu_{3}\right)  \Gamma\left(  \nu_{4}\right)  \Gamma\left(  2-\nu_{3}-\nu
_{4}\right)  }. \tag{4.20}%
\end{align}

It should be noted that the four-point conformal blocks for different channels
are related by the Racah coefficients \cite{ism07}, \cite{der19} for principal
series of representations of $SL(2,C)$.

\appendix

\section{Decomposition of tensor products of scalar principal series
representations}

The problem of decomposing the tensor product of two irreducible
representations of the group $SL(2,C)$ into irreducible ones was completely
solved by Naimark in \cite{naim1}. Here we give a brief overview of Naimark's
results on the decomposition of tensor products of scalar principal series representations.

The tensor product $T^{\sigma_{1}}\otimes T^{\sigma_{2}}$ can be realized in
the Hilbert space $L^{2}(C^{2})$ of functions $f\left(  z_{1},z_{2}\right)  $
of two variables $z_{1}$and $z_{2}$%
\begin{align}
&  \left[  T^{\sigma_{1}}\otimes T^{\sigma_{2}}\left(  g\right)  f\right]
\left(  z_{1},z_{2}\right)  \nonumber\\
&  =\left\vert g_{12}z_{1}+g_{22}\right\vert ^{2\sigma_{1}}\left\vert
g_{12}z_{2}+g_{22}\right\vert ^{2\sigma_{2}}f\left(  \frac{g_{11}z_{1}+g_{21}%
}{g_{12}z_{1}+g_{22}},\frac{g_{11}z_{2}+g_{21}}{g_{12}z_{2}+g_{22}}\right)
.\tag{A.1}%
\end{align}
This representation is decomposed only into principal series representations
$T^{\chi}$, and into those and only those for which $s$ is an integer. The
components of this expansion, transforming according to the representation
$T^{\chi}$, are determined by the formula%
\begin{equation}
f\left(  z\chi\right)  =\int_{C^{2}}N\left(  z_{1}\sigma_{1}z_{2}\sigma
_{2};z\chi\right)  f\left(  z_{1},z_{2}\right)  dz_{1}dz_{2},\tag{A.2}%
\end{equation}
where%
\begin{align}
&  N\left(  z_{1}\sigma_{1}z_{2}\sigma_{2};z\chi\right)  \nonumber\\
&  =\left\vert z_{2}-z_{1}\right\vert ^{s-\sigma-\sigma_{1}-\sigma_{2}%
-4}\left(  z_{2}-z_{1}\right)  ^{-s}\left\vert z-z_{1}\right\vert
^{-s+\sigma-\sigma_{1}+\sigma_{2}}\left(  z-z_{1}\right)  ^{s}\nonumber\\
&  \times\left\vert z_{2}-z\right\vert ^{-s+\sigma+\sigma_{1}-\sigma_{2}%
}\left(  z_{2}-z\right)  ^{s}.\tag{A.3}%
\end{align}
The function $f\left(  z_{1},z_{2}\right)  $ is expressed through these
components by the formula%
\begin{equation}
f\left(  z_{1},z_{2}\right)  =\frac{1}{4\pi^{4}}%
%TCIMACRO{\dsum \limits_{s=-\infty}^{\infty}}%
%BeginExpansion
{\displaystyle\sum\limits_{s=-\infty}^{\infty}}
%EndExpansion%
%TCIMACRO{\dint \limits_{-\infty}^{\infty}}%
%BeginExpansion
{\displaystyle\int\limits_{-\infty}^{\infty}}
%EndExpansion
\left(  s^{2}+\rho^{2}\right)  d\rho\int\overline{N\left(  z_{1}\sigma
_{1}z_{2}\sigma_{2};z\chi\right)  }f\left(  z\chi\right)  dz.\tag{A.4}%
\end{equation}
In this case, the Plancherel formula is%
\begin{equation}
\int_{C^{2}}\left\vert f\left(  z_{1},z_{2}\right)  \right\vert ^{2}%
dz_{1}dz_{2}=\frac{1}{4\pi^{4}}%
%TCIMACRO{\dsum \limits_{s=-\infty}^{\infty}}%
%BeginExpansion
{\displaystyle\sum\limits_{s=-\infty}^{\infty}}
%EndExpansion%
%TCIMACRO{\dint \limits_{-\infty}^{\infty}}%
%BeginExpansion
{\displaystyle\int\limits_{-\infty}^{\infty}}
%EndExpansion
\left(  s^{2}+\rho^{2}\right)  d\rho\int_{C}\left\vert f\left(  z\chi\right)
\right\vert ^{2}dz.\tag{A.5}\label{pl}%
\end{equation}
\qquad\qquad

It should be noted that the kernel $N$ is uniquely determined up to a constant
factor $C(\sigma_{1},\sigma_{2},\chi)$. Plancherel's formula (\ref{pl}) fixes
the absolute value of this factor. The phase of $C(\sigma_{1},\sigma_{2}%
,\chi)$ is arbitrary.

\end{document}